\documentclass[AMA,STIX1COL]{WileyNJD-v2}

\usepackage{comment}
\usepackage{subfigure}
\usepackage{dblfloatfix}
\usepackage{graphicx}

\newcommand{\Fig}[1]{Fig.~\ref{fig:#1}}

\newcommand{\Sec}[1]{Sec.~\ref{sec:#1}}

\newcommand{\Eq}[1]{(\ref{eq:#1})}

\articletype{Regular paper}%

\begin{document}

\title{Characterizing the Power Cost of Virtualization Environments}

\author[1]{Senay Semu Tadesse$^1$}

\author[1]{Carla Fabiana Chiasserini$^1$}

\author[1]{Francesco Malandrino$^1$}

\address[1]{\orgdiv{DET}, \orgname{Politecnico di Torino}, \orgaddress{\state{TO}, \country{Italy}}}

\corres{Francesco Malandrino\\\email{francesco.malandrino@polito.it}}

\abstract[Summary]{
Virtualization is a key building block of next-generation mobile networks. It can be implemented through two main approaches: traditional virtual machines and lighter-weight containers. Our objective in this paper is to compare these approaches and study the power consumption they are associated with. To this end, we perform a large set of real-world measurements, using both synthetic workloads and real-world applications, and use them to model the relationship between the resource usage of the hosted application and the power consumption of both virtual machines and containers hosting it. We find that containers incur substantially lower power consumption than virtual machines, and that such consumption increases more slowly with the application load.
}



\maketitle

\section{Introduction}
\label{sec:intro}

One of the most notable differences between next-generation (``5G'') mobile networks and present-day ones is the former's ability to serve user request within the network itself. Instead of traveling all the way to an Internet-based server, requests will be processed at computation-capable nodes belonging to the backhaul network itself. This paradigm is known as multi-access edge computing (MEC)~\cite{fog,mec}.

Virtualization is a key enabling technology of MEC, as it allows general-purpose network hardware to process requests of any type. At the same time, it is associated with an overhead in terms of CPU usage and memory occupation; such overhead in turn translates into additional power consumption. Power consumption is an increasingly important key performance indicator (KPI) for all types of networks, as it affects their profitability as well as their environmental sustainability. In this context, it is important that the virtualization technique is chosen accounting for the associated power consumption.

\begin{figure}
\centering
\includegraphics[width=0.6\textwidth]{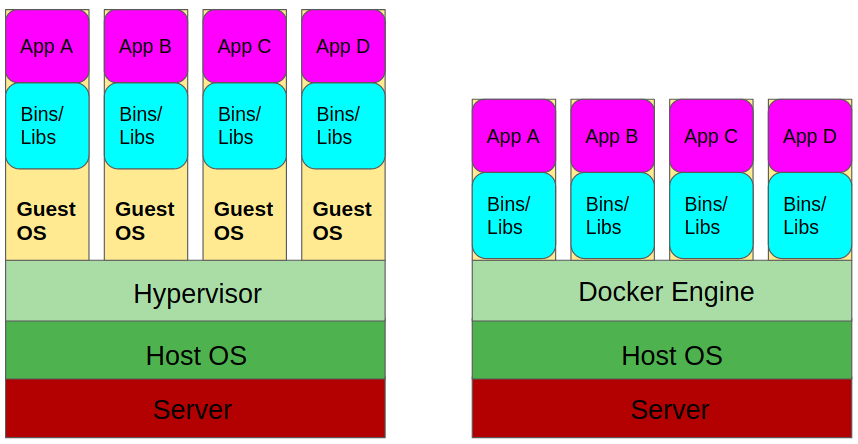}
\caption{
High-level architecture of virtual machine-based virtualization (left) and container-based virtualization (right).
\label{fig:comparison}
}
\end{figure}

The traditional approach is virtual machine (VM)-based virtualization~\cite{vms}: as summarized in \Fig{comparison}, a hypervisor software emulates a whole virtual machine, running the guest operating system and applications. The main advantage of VM-based virtualization is the high level of separation between guests and host: guest machines can have different architecture (e.g., x86 and ARM) and different operating system (e.g., Windows or Linux) from their host; furthermore, the same host can run multiple guests with different architectures and operating systems at the same time. On the negative side, the tasks of emulating guest hardware and running the guest operating system translate into a significant CPU and memory overhead.

More recently, container-based virtualization~\cite{containers} has emerged as a lighter-weight alternative to VMs. As shown in \Fig{comparison}, both the hardware and the operating system kernel are shared between host and guest applications. Isolation is obtained through operating system-level primitives such as namespaces and cgroups, controlling the parts of the execution environment (running processes, daemons...) and system resources visible to guest applications running within containers. The main advantage of container-based virtualization is its lower overhead compared to VMs. Main drawbacks include a lower level of isolation between host and guests (which can lead to security issues) and the impossibility of emulating a guest operating system different from the host one.

In this paper, we embark on the twofold task of (i) measuring the power consumption associated with VM- and container-based virtualization under a variety of workloads, and (ii) modeling how such consumption depends on the workload at hand. By fulfilling the first task, we can check to which extent the lower overhead touted by containers translates into lower power consumption. Perhaps more importantly, studying the link between workload, chosen virtualization technique and power consumption allows us to estimate the power consumption associated with a generic workload, beyond those we directly test.

The rest of this paper is organized as follows. We begin by describing our methodology in \Sec{methodology}, including our testbed and the workloads we use. Then, in \Sec{synthetic} and \Sec{real} we summarize the results we obtain from synthetic and real-world workloads respectively. Finally, after summarizing related work in \Sec{relwork}, we conclude the paper in \Sec{conclusion}.

\section{Methodology}
\label{sec:methodology}

We measure the power consumption associated with virtualization through a small-scale testbed. Whenever possible, we use commodity hardware and open-source software, so as to make our results as easy to reproduce as possible. This section describes the hardware (\Sec{hardware}) and software (\Sec{software}) we use in our testbed, as well as the workloads, both synthetic and real, we take into account (\Sec{workloads}).

\subsection{Hardware}
\label{sec:hardware}

The host machine we use for both containers and virtual machines is an HP EliteBook 820 G3 laptop, equipped with a Intel Core i5-6200U processor (two cores, 2.3~GHz, 3~MByte cache) and 8~GByte of RAM. When testing client-server applications, we run the servers (within containers or VMs) on the laptop, and the clients on a separate computer, namely, a ThinkCenter M93p desktop. By doing so, we prevent client and server applications from interfering with one another, e.g., triggering thread preemption.
The laptop and desktop are connected through a portable Gigabit Ethernet switch, and that switch is not shared with other computers. This ensures that the connection between clients and servers is consistently fast, and never represents a bottleneck for the application running.

Finally, we measure the power consumed by the laptop through an RCE PM600 power meter, displaying the total power the laptop draws from the grid. By using a power meter instead of estimates from the operating system, we obtain very precise and reliable measurements. However, this also means that special care is needed to tell the power consumption due to the workload apart from other contributions, e.g., the power consumed by the host operating system. To this end,
\begin{itemize}
\item before running any workload, we measure the idle power consumption of the computer;
\item we subtract such value from the power consumption measured under each workload.
\end{itemize}
Additionally, we remove the battery from the laptop to ensure that no power is drawn to charge it.

\subsection{Software}
\label{sec:software}

There are two main decisions we need to make concerning the software to use in our testbed: the operating system to use, and which VM- and container-based virtualization solutions to adopt.

The natural choice for the operating system is Linux, due to its broad support for all virtualization technologies. Among Linux distributions we chose Ubuntu, which offers the best balance between up-to-date packages, documentation, and overall system stability; specifically, we use the 16.10~LTS version, with kernel version 4.8.0-46-generic. As both the i5 processor and this kernel support hyper-threading, four threads can be run at the same time.

VM-based virtualization is a mature technology, with several available options, both commercial and open-source. We choose VirtualBox, on the grounds that it is the most popular among open-source ones. Both commercial (most notably, VMWare) and open-source (namely, QEMU) alternatives are sometimes reported to be marginally faster than VirtualBox; however, VirtualBox is more widely available than the other two solutions (VMWare requires a license, and QEMU -- branded a ``processor emulator'' -- is suited for narrower set of scenarios).

Choosing the reference container-based solution is easier. Docker is indeed the de facto standard, and arguably the primary reason why container-based virtualization attained its current level of popularity. It is an open-source application, and Linux support is its primary (though not exclusive) focus.

\subsection{Workloads}
\label{sec:workloads}

We measure the power consumption under two types of workloads:
\begin{itemize}
\item synthetic workloads, consuming a predetermined amount of CPU resources, memory, storage, or network capacity;
\item real-world applications, with servers  running into VMs or containers.
\end{itemize}

Through synthetic workloads, we can establish which among CPU usage, memory and storage consumption, and network traffic has the highest impact on power consumption. Real-world workloads, on the other hand, allow us to get a glimpse of how popular, present-day applications would perform in a virtualized environment, and the power consumption we can expect from them.

\subsubsection{Synthetic workloads}

Synthetic workloads aim at stressing a specific component of the system, e.g., CPU or memory, while using as little as possible of the other resources. As an example, a good CPU test will consume a predictable and configurable amount of CPU, while at the same time occupying a negligible amount of memory, storing no data on the hard disk, and generating no network traffic. Whenever possible, we use existing, off-the-shelf test applications as our workloads, as described next.

\noindent {\bf CPU test}. We use the Sysbench application, simulating a high CPU load by testing prime numbers. Prime number testing is a CPU-intensive task, which requires little memory and no reading/writing from disk or network. This implies that the Sysbench process will almost never be put to sleep pending I/O operations, thus using (unless preempted by other processes) virtually 100\% of a (logical) CPU core. Since Sysbench is a single-threaded application, it will not use multiple cores even when they are available.

\noindent {\bf Memory test}. Our objective here is to read and write a significant amount of memory, while at the same time using the CPU as little as possible. To this end, we write a customized Java application that allocates and then copies a large array, without initializing or writing anything into it.

\noindent {\bf Network test}. We use the iperf program to generate a predetermined quantity of TCP traffic from servers (running within VMs or containers) and clients, running on the host laptop. We keep the clients on the same machine as the servers in order not to use any networking hardware, and only measure the power consumption due to the processing at the hypervisor and kernel.

\noindent {\bf Disk test}. We use the Flexible IO (FIO) application to transfer a 10-GByte file using a configurable block size. By doing so, we are able to adjust the number of disk operations needed for the transfer -- the smaller the block size, the more disk operations --, and thus estimate their impact on the power consumption.

\begin{figure}
\centering
\includegraphics[width=0.6\textwidth]{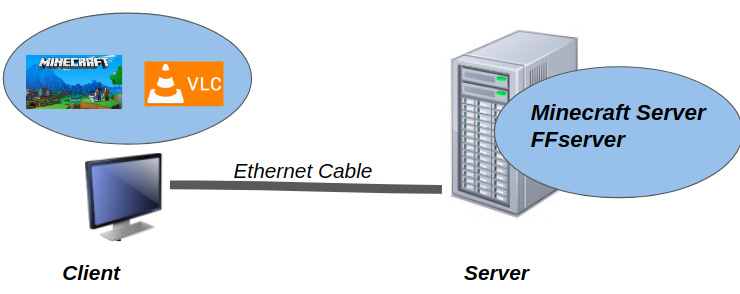}
\caption{
Testbed setup for real-world workloads.
\label{fig:realworld}
}
\end{figure}

\subsubsection{Real-world workloads}
\label{sec:sub-real}

For real-world workloads, we select two types of application that play a crucial role in the traffic of next-generation networks: video streaming and on-line gaming.

\noindent {\bf Video steaming}. We employ the FFServer open-source streaming server, and the VLC free client. A varying number of servers, each running in its own VM or container, stream a video to a varying number of clients, running on the desktop computer as standalone applications. This setup, depicted in \Fig{realworld}, allows us to assess how both the number of servers and the number of clients per server affect the power consumption.

\noindent {\bf On-line gaming}. We select the popular on-line game Minecraft, as it offers an open-source server. In selecting the client, we need to ensure that (i) it can run on a headless machine, i.e., without graphical interface, and (ii) we can reproduce multiple times the same input, i.e., the same game. To this end, we combine the Java-based Minecraft client with the xdtool keystroke emulator, both running on the desktop computer.

\section{Results and models: synthetic workloads}
\label{sec:synthetic}

\subsection{No workload}

Before we apply any workload, we measure the idle power consumption associated with VirtualBox and Docker. To this end, we create several VMs or containers, leave them idle, and measure the subsequent increase in CPU usage and power consumption.

\begin{figure}
\centering
\subfigure[]{
\includegraphics[width=0.35\textwidth]{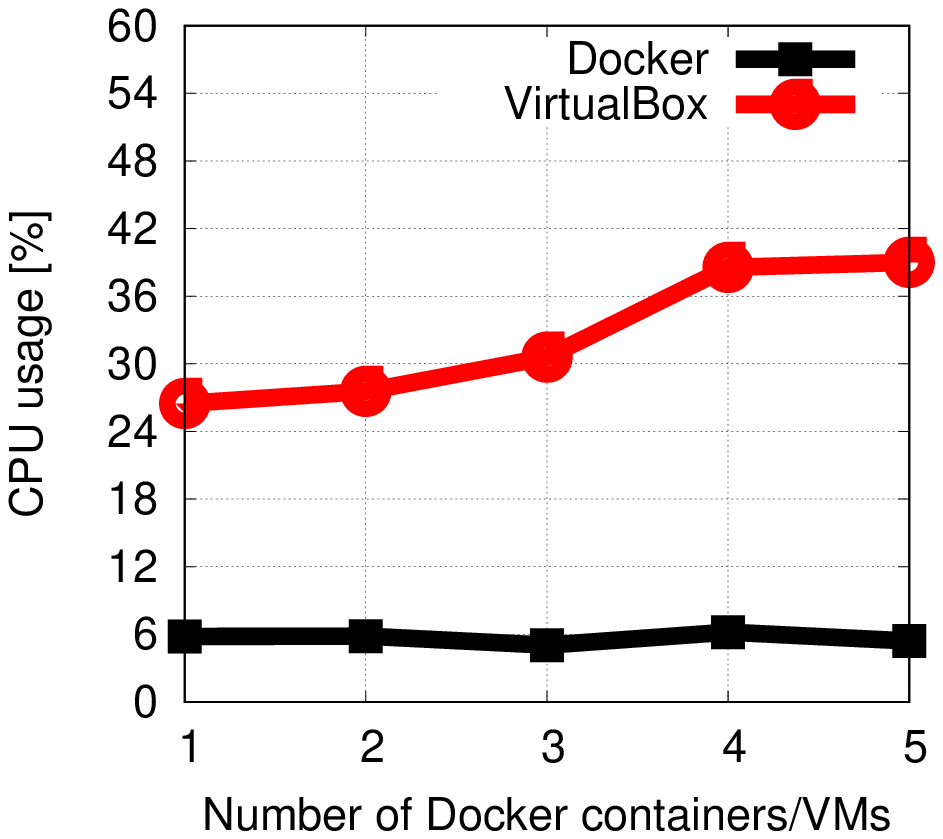}
} 
\subfigure[]{
\includegraphics[width=0.35\textwidth]{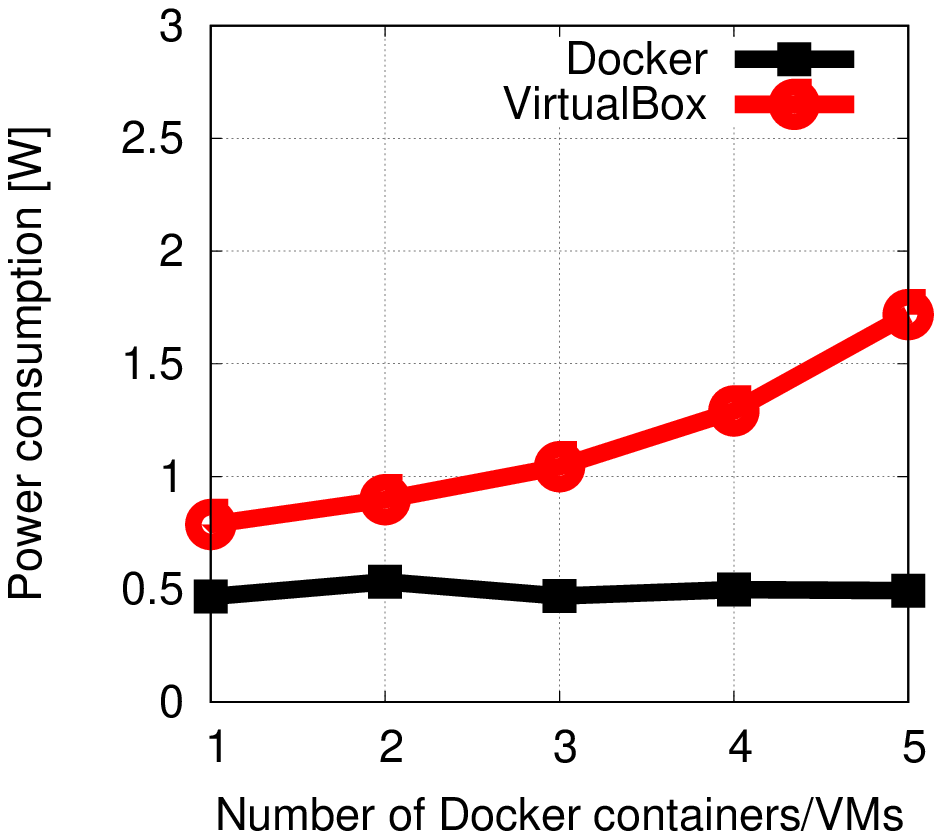}
} 
\caption{
No workload: CPU usage (a) and power consumption (b) as a function of the number of VMs/containers created, for VirtualBox (red lines) and Docker (black lines).
\label{fig:idle}
} 
\end{figure}

The results are summarized in \Fig{idle}, and already point at a fundamental difference between VMs and containers. Not only Docker is associated with a lower CPU usage -- and, thus, a lower power consumption -- than VirtualBox, but its overhead remains constant as the number of containers increases. The overhead associated with VirtualBox, on the other hand, grows with the number of idle VMs we create: we can identify an offset, due to running the hypervisor, and a linear growth as the number of VMs (and thus the quantity of virtual hardware to emulate) increases. Intuitively, \Fig{idle} also suggests that there is a penalty associated with creating a VM and then leaving it unused, while containers only consume system resources if the application they host is active.

\subsection{CPU- and memory-intensive workloads}

\begin{figure}
\centering
\subfigure[\label{fig:cpu-cpu} ]{
\includegraphics[width=0.32\textwidth]{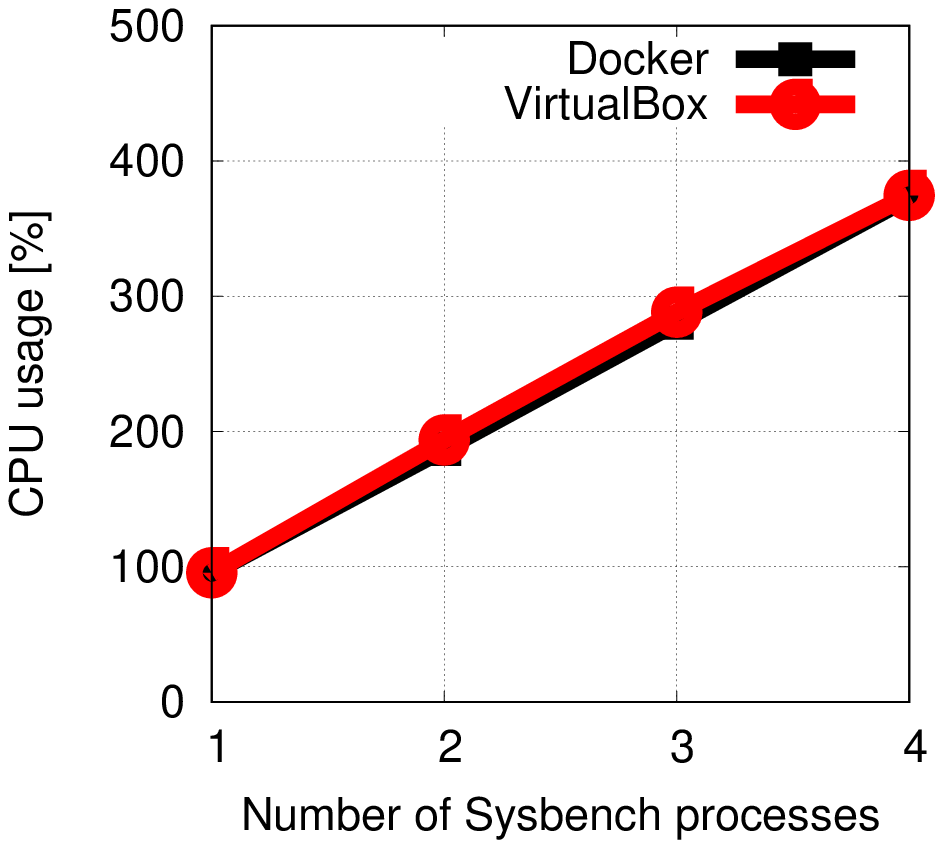}
} 
\subfigure[\label{fig:cpu-power} ]{
\includegraphics[width=0.32\textwidth]{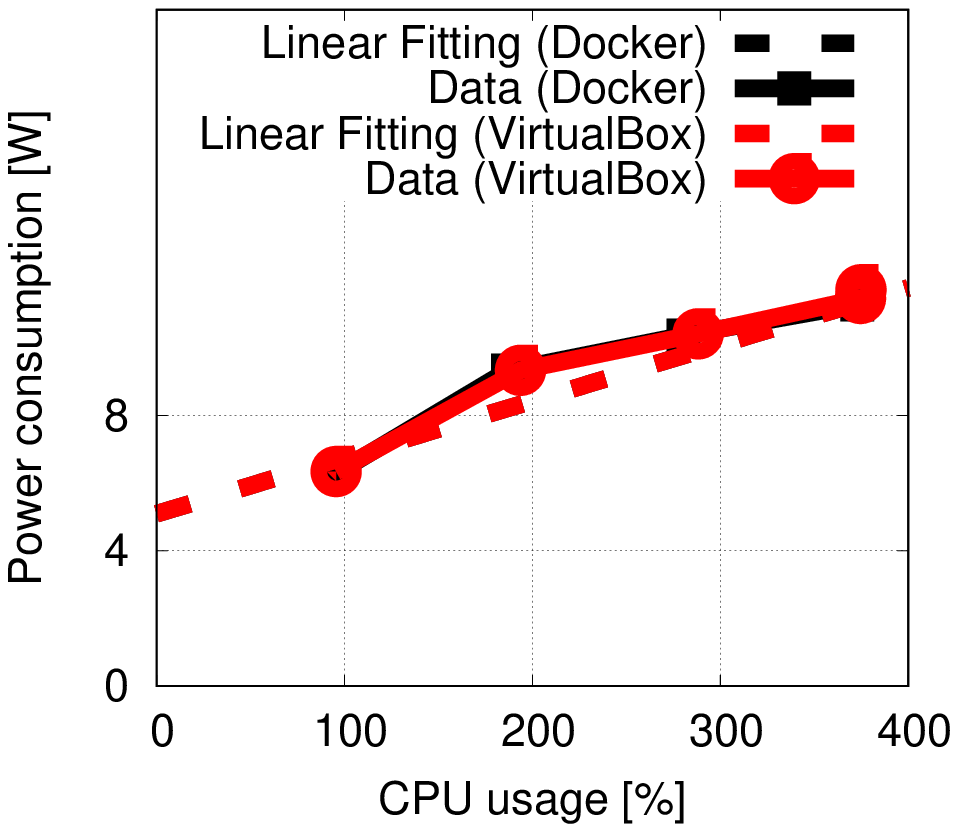}
} 
\subfigure[\label{fig:cpu-work} ]{
\includegraphics[width=0.32\textwidth]{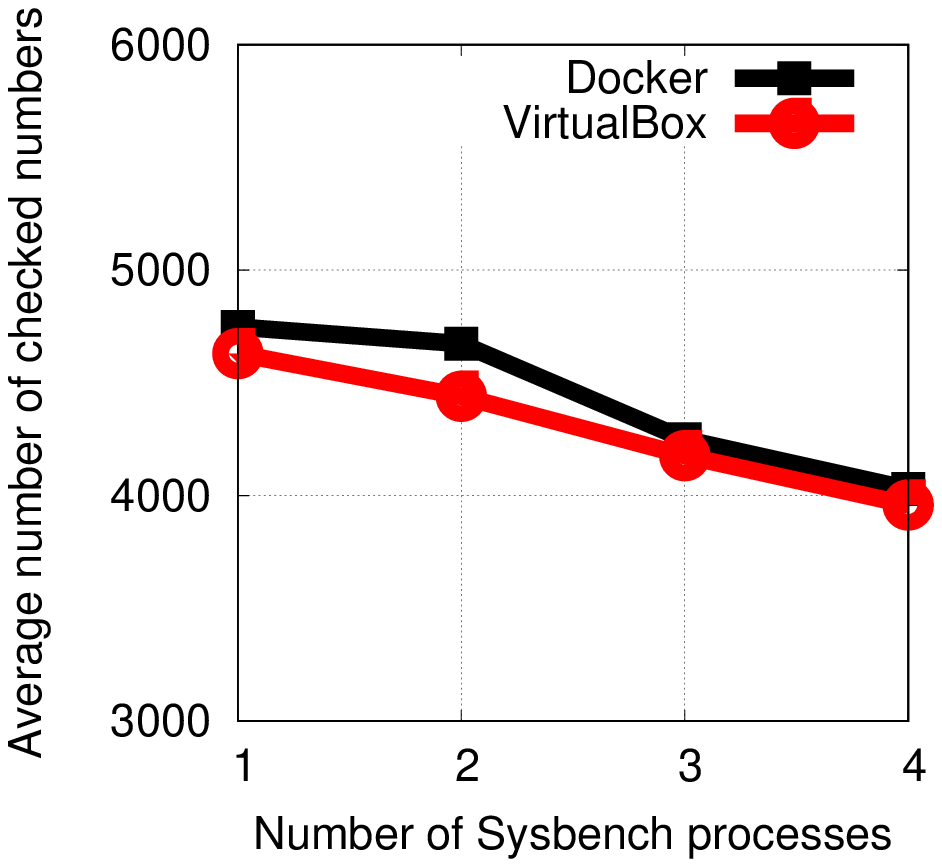}
} 
\caption{
CPU-intensive workload: CPU usage (a), and power consumption (b) and work performed (c) as a function of the number of Sysbench processes, for VirtualBox (red lines) and Docker (black lines).
\label{fig:cpu}
} 
\end{figure}

We now move to the CPU-intensive workload and show, in \Fig{cpu}, how the number of Sysbench processes we start influences the power consumption.

\begin{figure}
\centering
\includegraphics[width=0.35\textwidth]{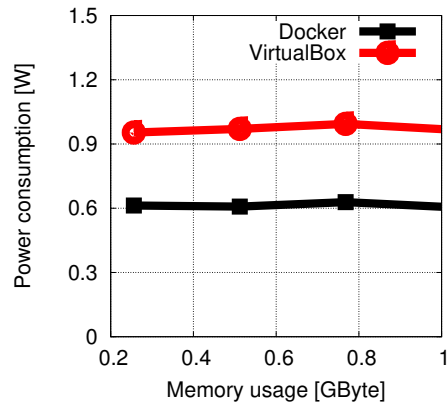}
\caption{
Memory-intensive workload: Power consumption as a function of the quantity of memory copied, for VirtualBox (red lines) and Docker (black lines).
\label{fig:mem}
} 
\end{figure}

\Fig{cpu-cpu} shows that the amount of CPU usage grows linearly with the number of Sysbench processes we start. Recall that Sysbench is a single-threaded application, so if (for example) we launch two Sysbench processes then exactly two CPU cores will be used. \Fig{cpu-power}, more interestingly, shows that the power consumption also grows linearly with the CPU usage, which suggests that CPU usage is its main component. One might be surprised to see almost no difference between VirtualBox and Docker curves in \Fig{cpu-cpu} and \Fig{cpu-power}. We observed from \Fig{idle} that Docker has a lower overhead than VirtualBox, so we could expect a lower CPU usage in \Fig{cpu-cpu} and thus a lower power consumption in \Fig{cpu-power}.

The solution to this apparent paradox is given in \Fig{cpu-work}, showing the quantity of work made by Sysbench, i.e., how many numbers it can check: using Docker instead of VirtualBox means that Sysbench is able to do more work in the same time period. In summary, the Sysbench test uses the same amount of CPU when VirtualBox and Docker are used; however, in the VirtualBox case, a higher fraction of that CPU is consumed by virtualization overhead, which results in less CPU available for actual work.

\Fig{mem} depicts the power consumption resulting from our Java-based memory test. We can clearly see that the amount of memory we copy has almost no impact on the total power consumption. Indeed, memory operations require very little intervention from the CPU; \Fig{mem} also confirms our intuition that CPU usage is the main contribution to the total memory consumption.

\subsubsection{Linear fitting}
\label{sec:sub-cpumodel}

We use the data summarized in \Fig{cpu-power} to extrapolate a linear model linking CPU usage with power consumption. Specifically, we use the ordinary least squares~\cite{ols} method to find the linear relationship that minimizes the sum of squared errors. The fitted relationship for VirtualBox is the following:
\begin{equation}\label{eq:cpuvm}
P_\text{[W]}=0.0175c_\text{[cores]}+4.9091.
\end{equation}
For Docker, we have the following:
\begin{equation}\label{eq:cpudocker}
P_\text{[W]}=0.0166c_\text{[cores]}+5.0444.
\end{equation}
In both \Eq{cpuvm} and \Eq{cpudocker}, $P$~is the power consumption (in watt), and $c$~represents the CPU usage in cores.
The average fitting error is very low, as little as~0.18\% for VirtualBox and 0.04\% for Docker. This confirms our intuition that CPU usage is the main driver of power consumption, and the link between them is as simple as a linear relationship.

\subsection{Network-intensive workloads}

We now use iperf to create a network-intensive workload. Specifically, we use a single server (running in a container or VM) and a single client (running directly on the laptop), and change the offered data rate between~0.01 and~0.9 GBit/s.

\begin{figure}
\centering
\subfigure[\label{fig:net1-power}]{
\includegraphics[width=0.35\textwidth]{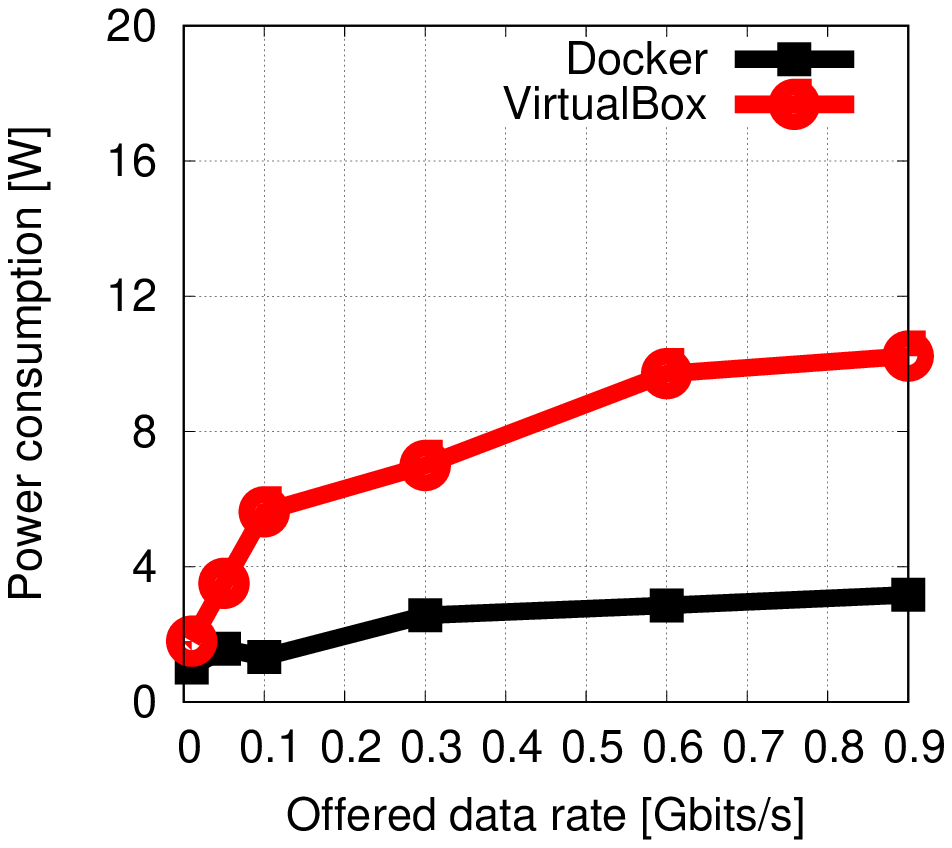}
} 
\subfigure[\label{fig:net1-cpu}]{
\includegraphics[width=0.35\textwidth]{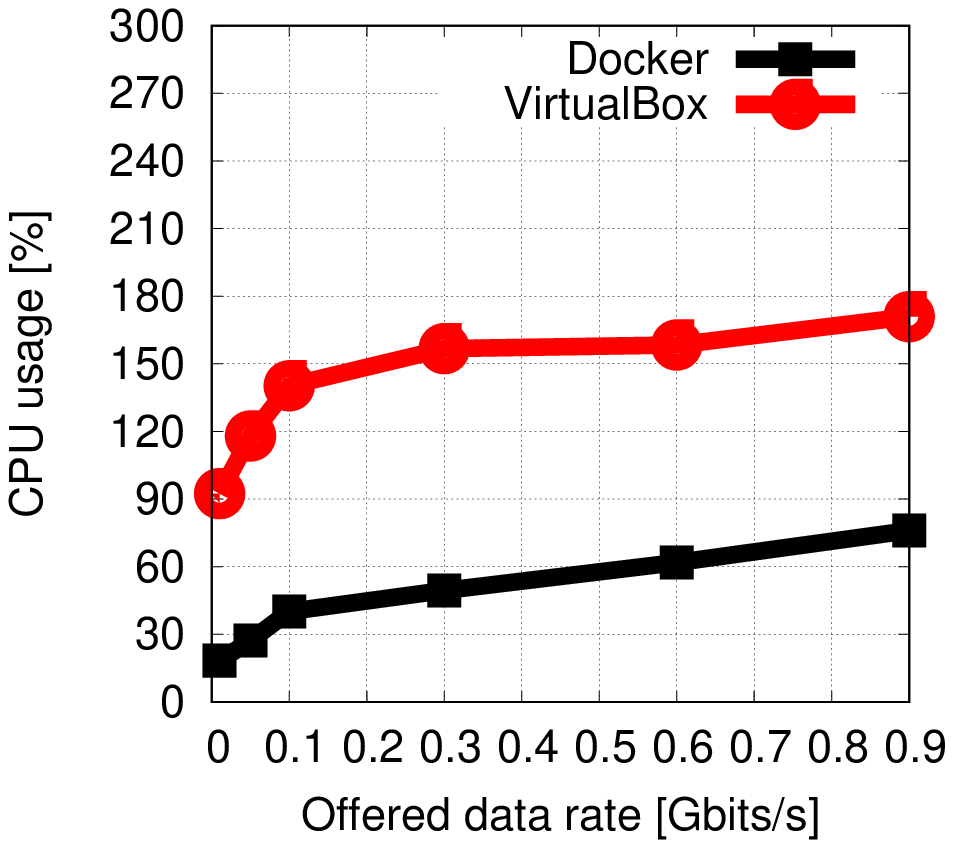}
} 
\caption{
Network-intensive workload with one client and one server: power consumption (a) and CPU usage (b) as a function of the offered traffic, for VirtualBox (red lines) and Docker (black lines).
\label{fig:net1}
} 
\end{figure}

\Fig{net1-power} depicts the power consumption as a function of the offered traffic, and highlights several very important facts. First, VirtualBox exhibits a higher power consumption than Docker, consistently with what we observed earlier. Even more importantly, such power consumption grows much faster for VirtualBox than for Docker, suggesting that containers have not only a better efficiency than VMs, but also a better scalability.

The reason for this difference lies in the operations needed to transfer packets between the iperf client and server. In Docker, all that is needed is that packets traverse the networking stack of the host operating system's kernel. In VirtualBox, packets need to traverse the host kernel, the hypervisor, and then the guest kernel. As shown in \Fig{net1-cpu}, all these operations take a substantial toll on the CPU usage, which in turn translates into the higher power consumption we observed in \Fig{net1-power}.


\subsection{Disk-intensive workload}

\begin{figure}
\centering
\subfigure[\label{fig:disk-power}]{
\includegraphics[width=0.35\textwidth]{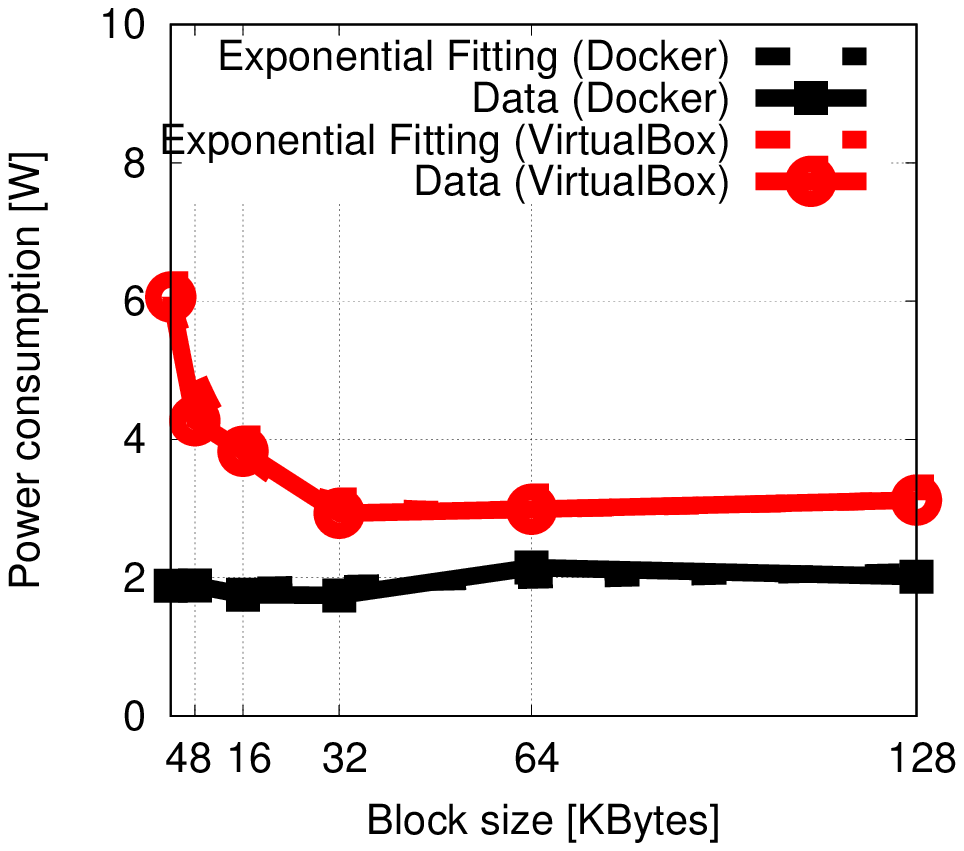}
} 
\subfigure[\label{fig:disk-latency}]{
\includegraphics[width=0.35\textwidth]{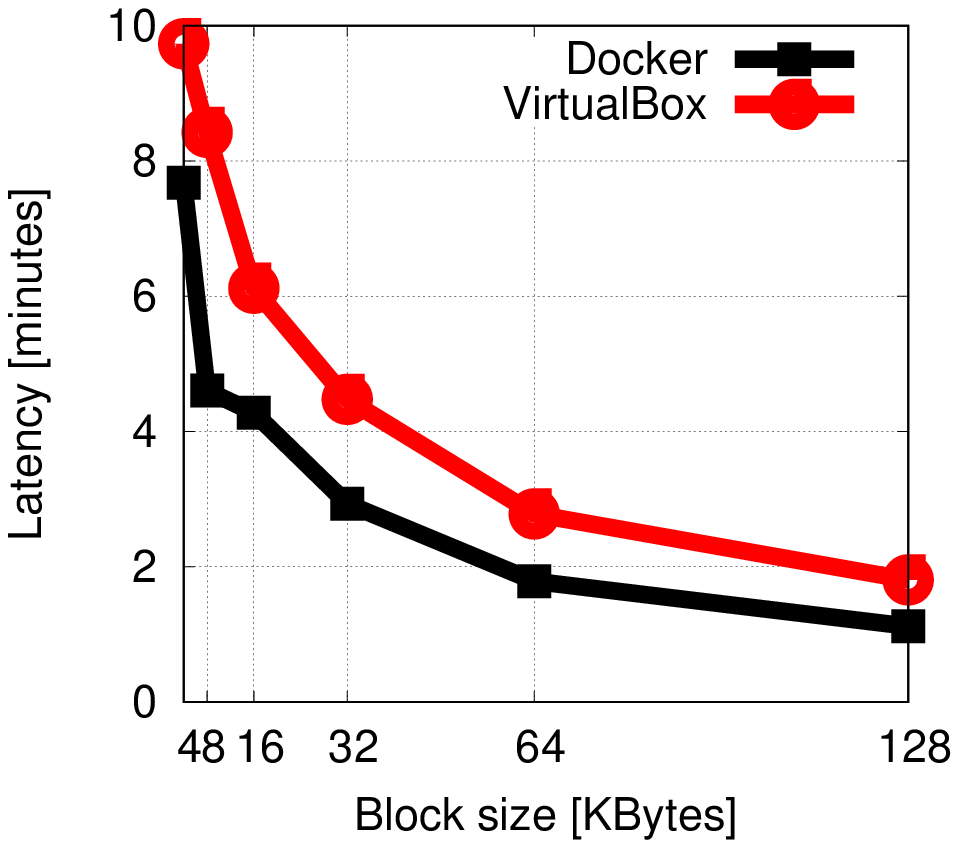}
} 
\caption{
Disk-intensive workload: power consumption (a) and transfer time (b) as a function of the block size, for VirtualBox (red lines) and Docker (black lines).
\label{fig:disk}
} 
\end{figure}

As described earlier, our disk-intensive workload consists in copying a 10-GByte file with different block sizes; the smaller the block size, the larger the number of I/O operations that are required. \Fig{disk-power} summarizes the power consumption resulting from such workload for different block sizes. We can observe that containers always exhibit a low power consumption, consistently with previous tests. More importantly, the power consumption of Docker is almost constant with respect to the block size, while the one of VirtualBox exhibits a steep increase as the block size drops below 32~kByte. The reason is similar to the one discussed for the network-intensive workload: a disk operation in VirtualBox requires the hypervisor to simulate the behavior of the hardware, which implies substantial CPU overhead and thus higher power consumption.

\Fig{disk-latency} shows time elapsed while performing the disk transfer. We can observe that, for both Docker and VMs, smaller block sizes translate into higher transfer times. However, transfer times with Docker are consistently shorter than transfer times with VirtualBox: the higher overhead incurred by virtual machine translates not only in higher power consumption but also in longer transfer times.

\subsubsection{Exponential fitting}

We use the data summarized in \Fig{disk-power} to extrapolate an exponential model linking block size with power consumption. As in the linear case, we use the least-squares method~\cite{ols} to find the exponential relatioship that minimizes the sum of squared errors. The fitted relationship for VirtualBox is the following:
\begin{equation}
\label{eq:diskvm}
P_\text{[W]}=4.81\exp\left(-0.1014b_\text{kByte}\right)+2.848\exp\left(0.00072b_\text{kByte}\right).
\end{equation}
For Docker, we have:
\begin{equation}
\label{eq:diskdocker}
P_\text{[W]}=1.857\exp\left(0.0009083b_\text{kByte}\right).
\end{equation}
In both \Eq{diskvm} and \Eq{diskdocker}, $P$~is the power consumption (in watt), and $b$~represents the block size kilobytes.
The average fitting error is 11.4\% for VirtualBox and 13.9\% for Docker. Such figures are higher than those for the linear model described in \Sec{sub-cpumodel}; this suggests a fairly complex relationship between block size and power consumption.

\section{Results and models: real-world workloads}
\label{sec:real}

We now move to the real-word workloads described in \Sec{sub-real}, i.e., the FFServer streaming application and the Minecraft game. As summarized in \Fig{realworld}, we run the servers on our laptop, each in its own VM or container, and the clients on a separate desktop computer.

\subsection{Video streaming: FFServer}

\begin{figure}
\centering
\subfigure[\label{fig:ff1-power}]{
\includegraphics[width=0.32\textwidth]{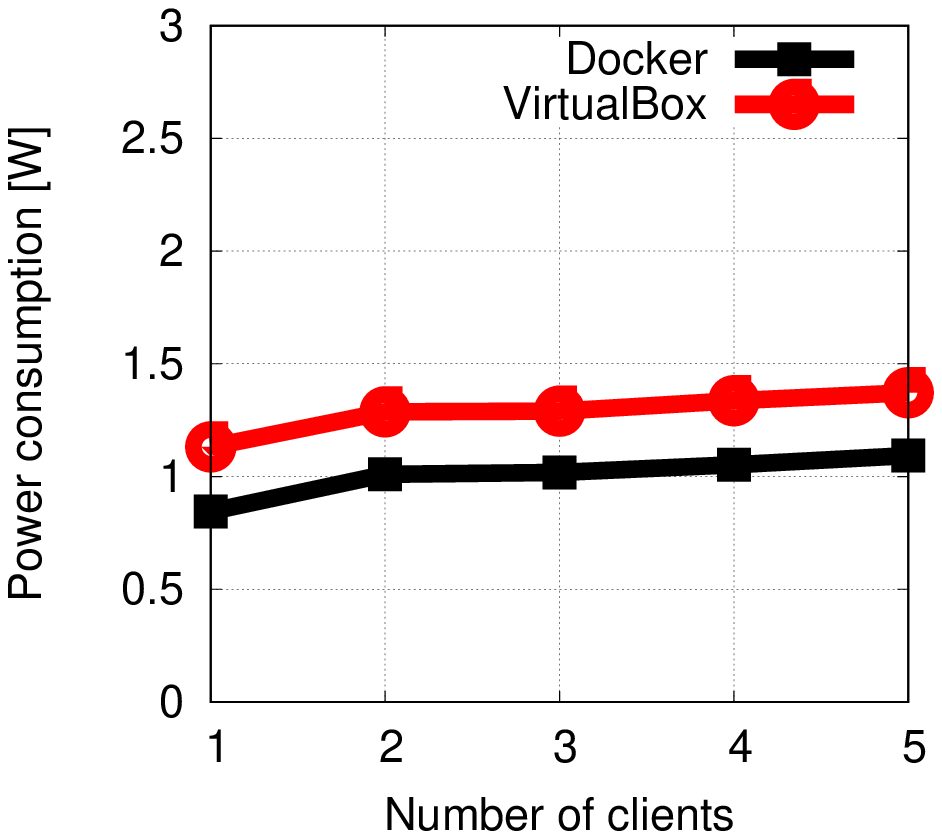}
} 
\subfigure[\label{fig:ff1-cpu}]{
\includegraphics[width=0.32\textwidth]{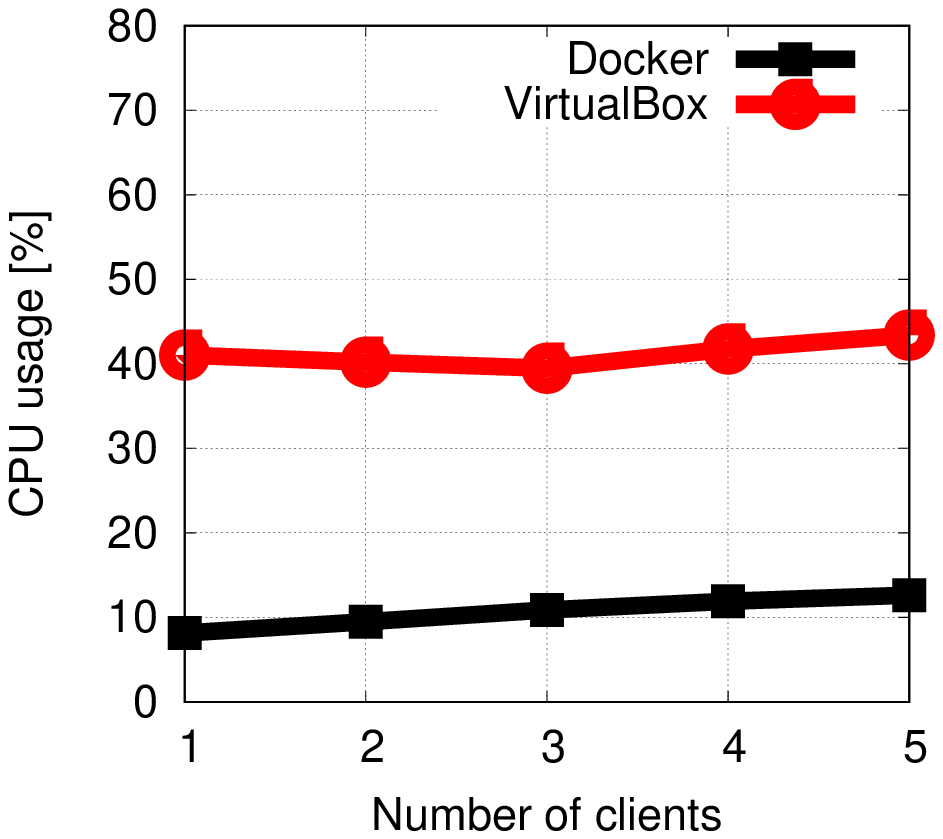}
} 
\subfigure[\label{fig:ff1-memory}]{
\includegraphics[width=0.32\textwidth]{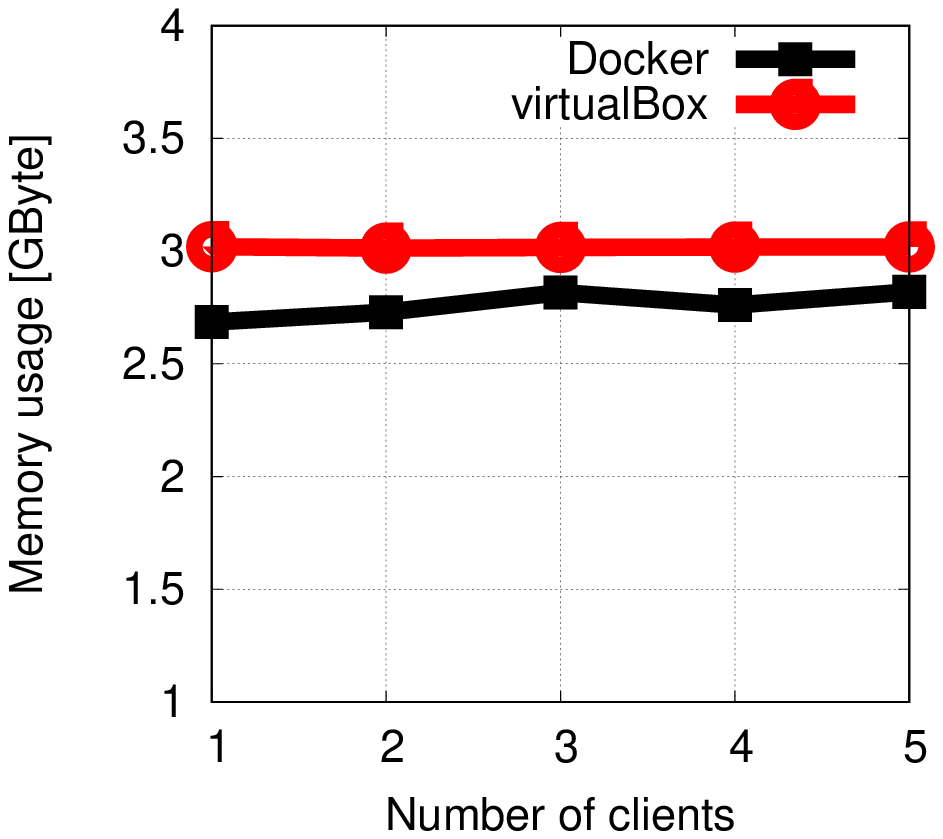}
} 
\caption{FFServer, single-server setup: power consumption (a), CPU usage (b) and memory usage (c) as a function of the number of clients, for VirtualBox (red lines) and Docker (black lines).
\label{fig:ff1}
} 
\centering
\subfigure[\label{fig:ffm-power}]{
\includegraphics[width=0.32\textwidth]{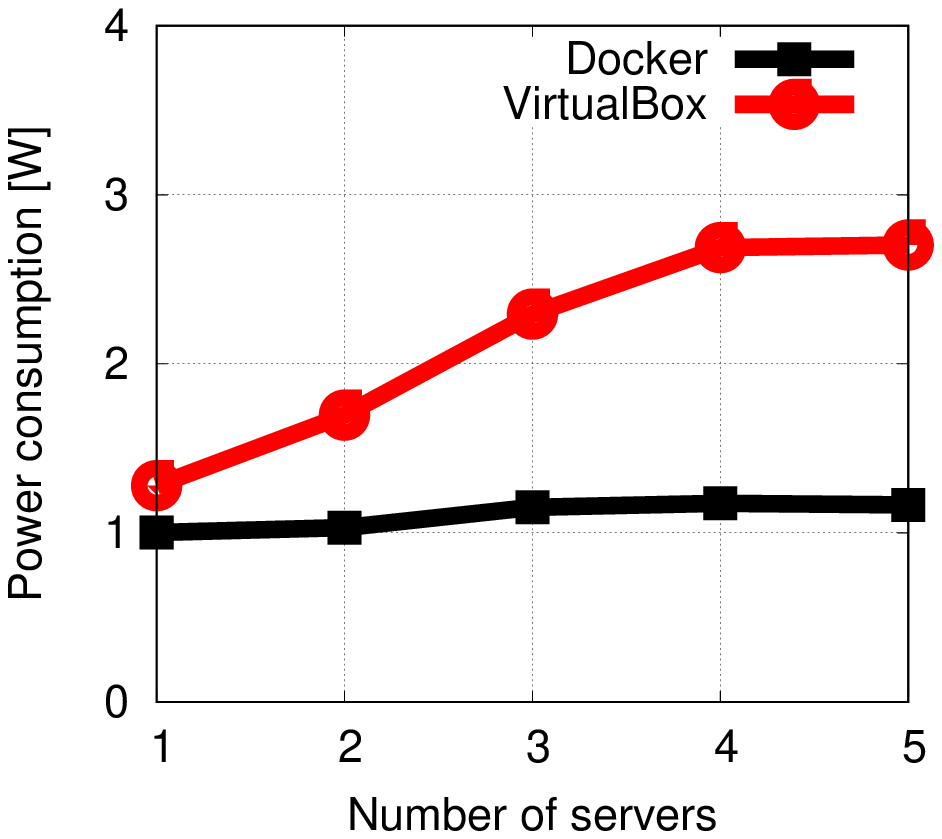}
} 
\subfigure[\label{fig:ffm-cpu}]{
\includegraphics[width=0.32\textwidth]{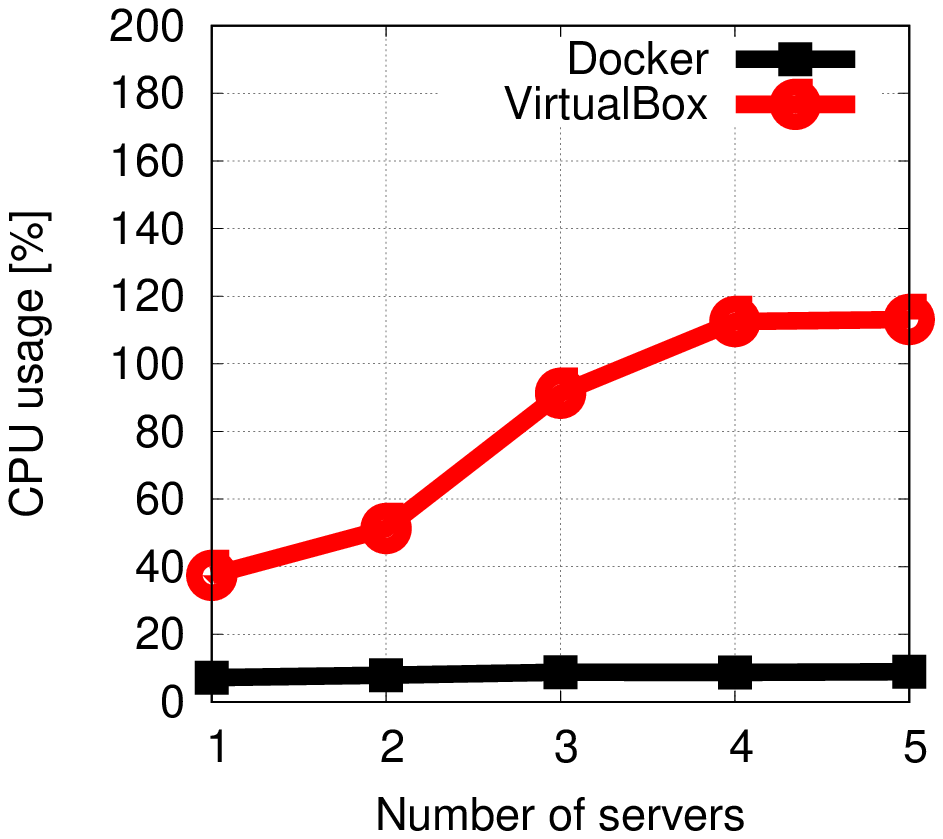}
} 
\subfigure[\label{fig:ffm-memory}]{
\includegraphics[width=0.32\textwidth]{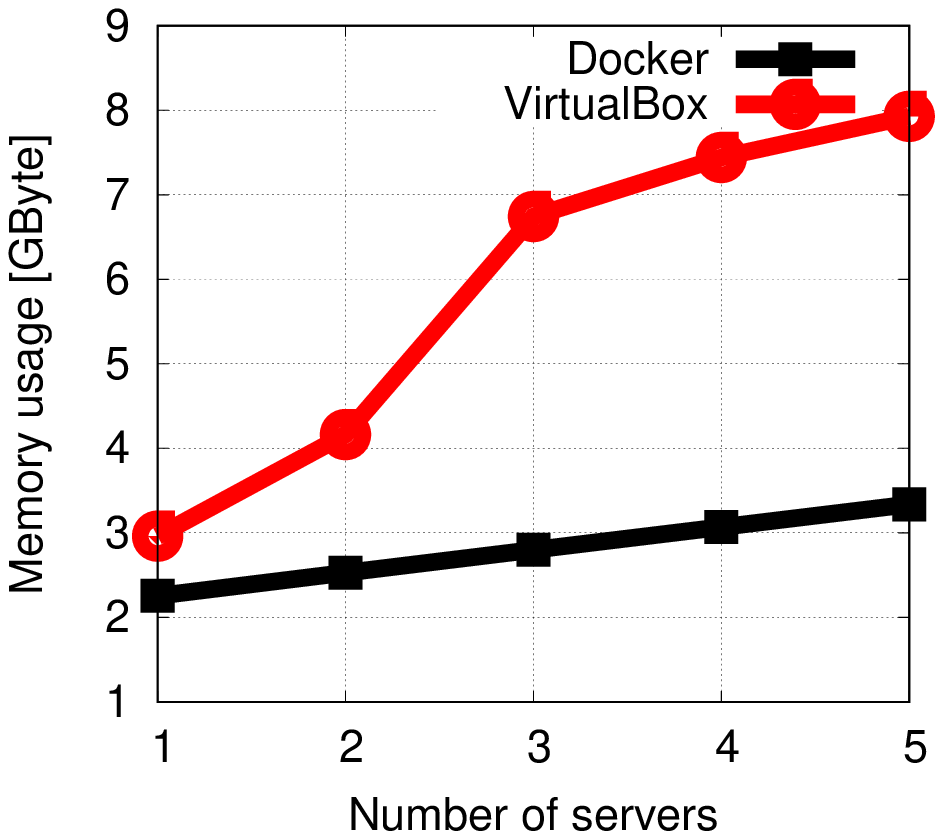}
} 
\caption{FFServer, multiple-server setup: power consumption (a), CPU usage (b) and memory usage (c) as a function of the number of clients, for VirtualBox (red lines) and Docker (black lines).
\label{fig:ffm}
} 
\end{figure}

In our first test, we deploy one FFServer server on our laptop (running within a VM or container) and use it to stream the same video to a varying number of VLC clients. It is important to highlight that we are modeling on demand streaming (\`{a} la YouTube) as opposed to real-time streaming (\`{a} la AceStream): each client plays an independent stream, and all traffic is unicast.

None the less, as we can see from \Fig{ff1}, the number of clients only has a small impact on the resources consumed by the server; indeed, neither the CPU usage (\Fig{ff1-cpu}) nor the memory consumption (\Fig{ff1-memory}) substantially increases with the number of clients. As a consequence, the power consumption (\Fig{ff1-power}) is essentially always the same.

We now study the effect of having multiple servers running in parallel on the laptop, each in its VM or container, and each serving one client. The corresponding resource consumption is shown in \Fig{ffm}.

Once more, we observe differences between Docker and VirtualBox that are not only quantitative, but qualitative. With Docker, five servers serving five clients consume approximately the same amount of CPU (\Fig{ffm-cpu}) and power (\Fig{ffm-power}) of one server serving five clients; only the memory usage shown in \Fig{ffm-memory} grows with the number of servers. Using VirtualBox, on the other hand, means that CPU usage and power consumption grow linearly with the number of servers.

We can conclude that the overhead incurred by container-based solutions like Docker is not only lower than that of VMs, but also grows more slowly with the number of containers being run. Such a scalability justifies the interest that container-based virtualization has attracted as a key technology of MEC, and indeed, one of its enablers.

\subsection{Gaming: Minecraft}

\begin{figure}
\centering
\subfigure[\label{fig:mc1-power}]{
\includegraphics[width=0.32\textwidth]{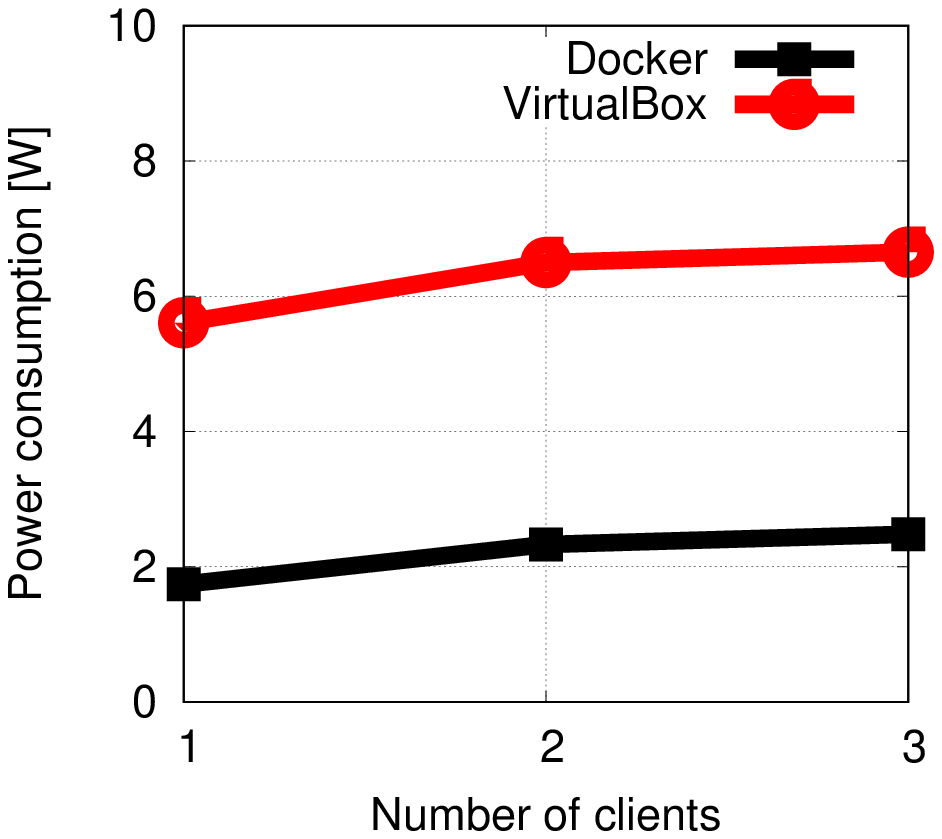}
} 
\subfigure[\label{fig:mc1-cpu}]{
\includegraphics[width=0.32\textwidth]{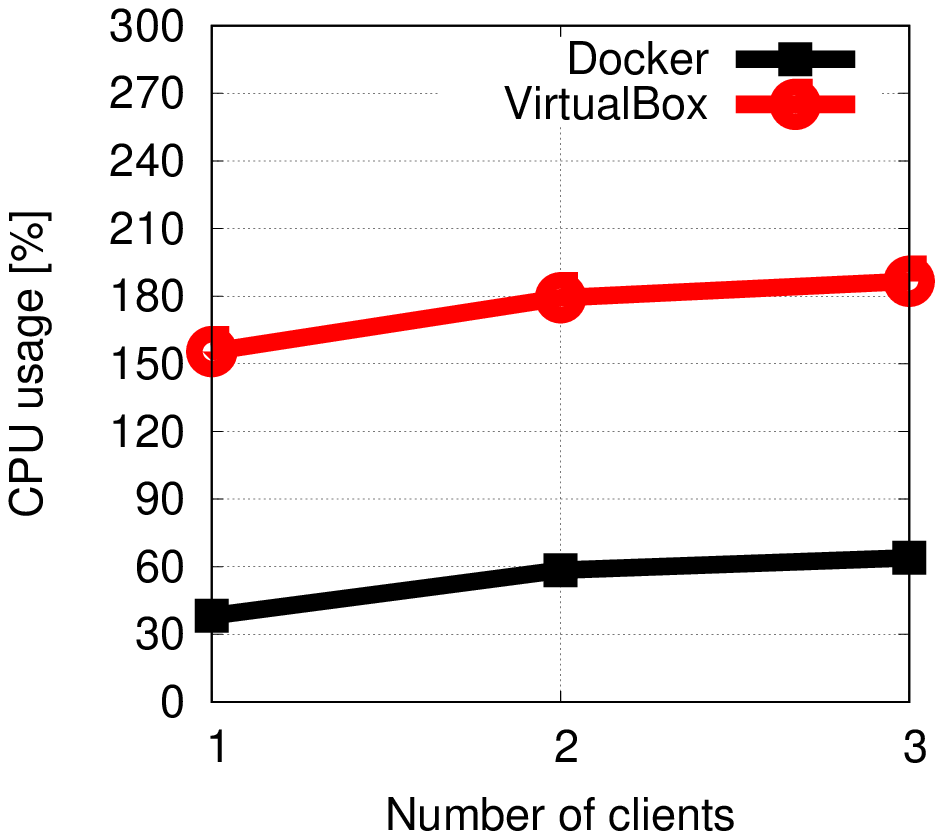}
} 
\subfigure[\label{fig:mc1-memory}]{
\includegraphics[width=0.32\textwidth]{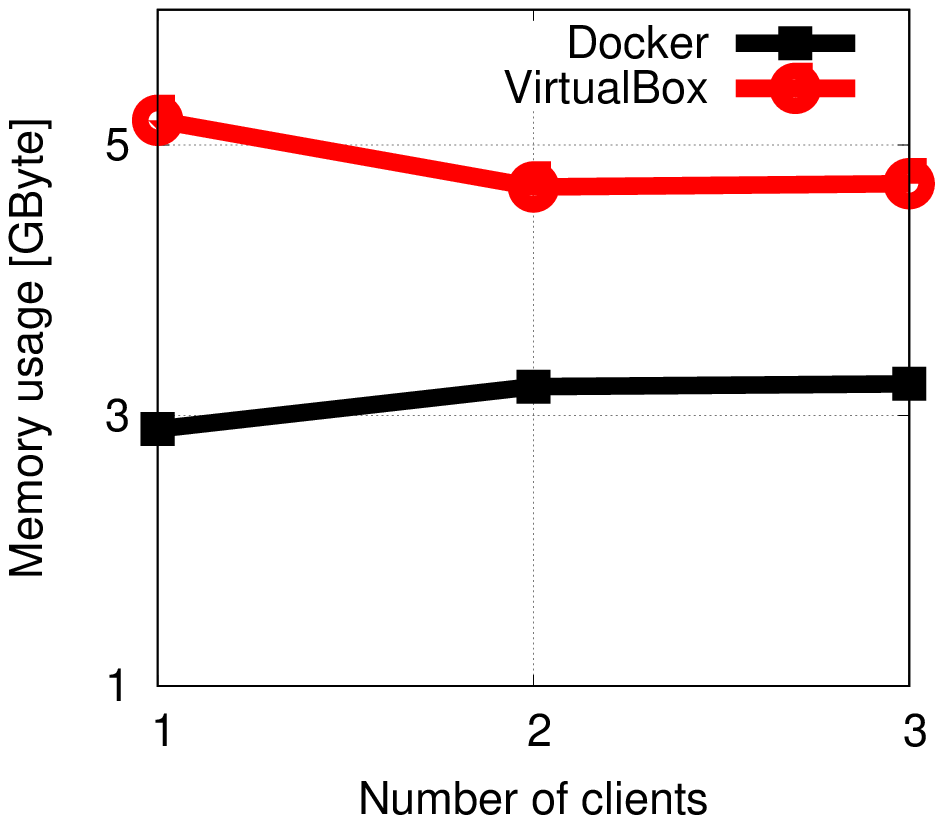}
} 
\caption{Minecraft, single-server setup: power consumption (a), CPU usage (b) and memory usage (c) as a function of the number of clients, for VirtualBox (red lines) and Docker (black lines).
\label{fig:mc1}
} 
\centering
\subfigure[\label{fig:mcm-power}]{
\includegraphics[width=0.32\textwidth]{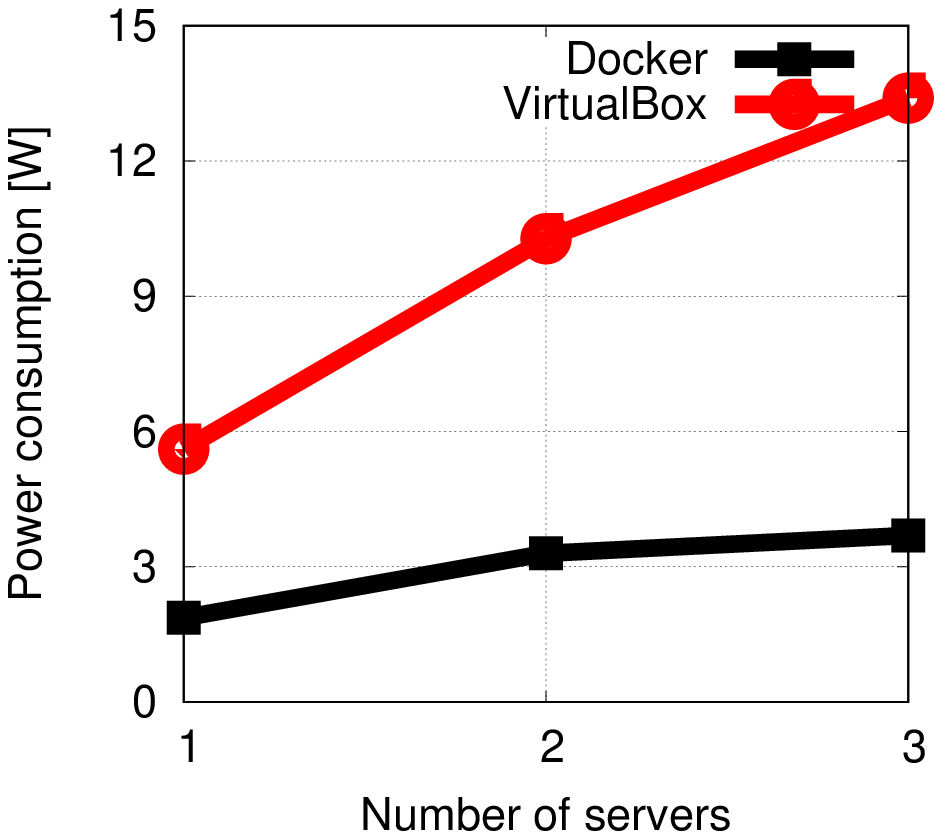}
} 
\subfigure[\label{fig:mcm-cpu}]{
\includegraphics[width=0.32\textwidth]{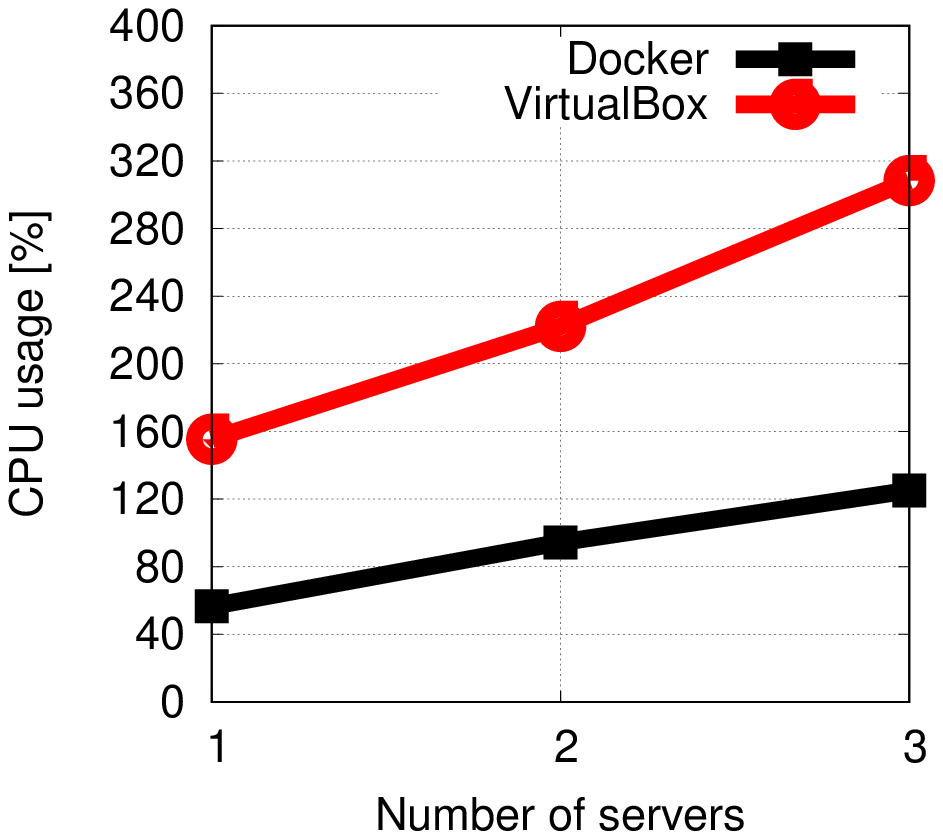}
} 
\subfigure[\label{fig:mcm-memory}]{
\includegraphics[width=0.32\textwidth]{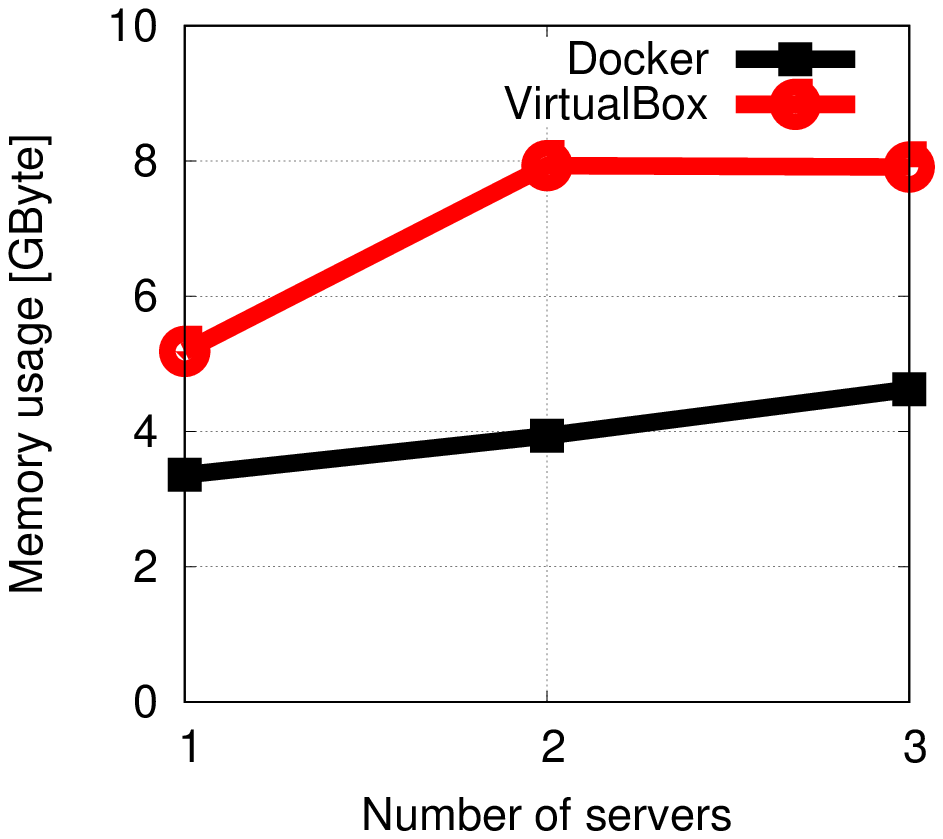}
} 
\caption{Minecraft, multiple-server setup: power consumption (a), CPU usage (b) and memory usage (c) as a function of the number of clients, for VirtualBox (red lines) and Docker (black lines).
\label{fig:mcm}
} 
\end{figure}

We now move to the Minecraft game. \Fig{mc1} refers to the setup with one server and a varying number of clients, similar to \Fig{ff1}. We can observe a steep increase in the utilization of resources as the number of clients increases, for both Docker and VirtualBox; this is due to the different nature of the application being run.

As far as the difference between VirtualBox and Docker is concerned, Docker still exhibits a substantially lower power consumption, (\Fig{mc1-power}), CPU usage (\Fig{mc1-cpu}), and memory usage (\Fig{mc1-memory}). It is also interesting to notice, from \Fig{mc1-cpu}, how the CPU overhead due to virtualization is higher than the load from the game server itself, intensive as it is.

\Fig{mcm} summarizes the resource consumption in the multi-server setup, when each user is served by its own server. Similar to \Fig{ffm}, we can see a higher resource consumption than in the single-server setup; furthermore, the qualitative behavior is now the same for both VirtualBox and Docker. The latter has, however, a much lower overhead than the former, suggesting that the scalability advantage of container-based virtualization is present under all types of workloads.

\section{Related work}
\label{sec:relwork}

Measuring the power consumption of virtualization environments and applications running on them is the subject of several existing works.
Kansal~et~al.~\cite{kansal} have developed Joulemeter, a tool to measure the energy consumption of a virtual machine and break it down as the sum of
the individual power consumptions of CPU, memory and disk.
Krishnan~et~al.~\cite{krishnan} have modeled the power consumption of virtual machines as a linear function of the number of CPU instructions and the
number of Last Level Cache (LLC) memory misses. Similar to what we obtained Kansal~et~al.~\cite{kansal} and Krishnan~et~al.~\cite{krishnan} has
obtained a linear relationship between the power consumption of the CPU intensive task and the number of instructions retired per second (CPU usage in
our case). The power consumption models obtained for memory intensive test of Krishnan~et~al.~\cite{krishnan} and  Kansal~et~al.~\cite{kansal} have
linear dependence on the last level cache misses.

Another VM power modelling technique is VMeter by Ata E ~et~al.\cite{ata}, which is based on online monitoring of CPU, cache, disk and RAM. The model
predicts instantaneous power consumption of an individual VM hosted on a physical node in addition to the full system power consumption with a model
which predicts the power consumption as the weighted sum of CPU, cache, disk and memory utilization. Yet another tool to measure power consumption of
virtualized applications are presented by Colmant ~et~al.\cite{maxime}. This paper \cite{maxime} presents a fine-grained monitoring middleware named
BITWATTS, which automatically learns an application-agnostic power model, which can be used to estimate the power consumption of applications.

Dhiman~et~al.~\cite{gaurav} present a system for online power prediction in virtualized environments based on Gaussian mixture models that use
architectural metrics of the physical and virtual machines (VM) collected dynamically by the system to predict both the physical machine and per VM
level power consumption. Bertran~\cite{bertran} proposed a system based on CPU and memory power models, relying on Performance Monitoring Counters
(PMCs), to perform energy accounting in virtualized systems. Bertran~\cite{bertran} have modeled the power consumption of VMs as the weighted sum of
performance counters by learning the weights of the model using incremental linear regression techniques.

Morabito~\cite{morabito} has presented an empirical investigation of different virtualization technologies (including containers) from the
viewpoint of power consumption. The work by Morabito~\cite{morabito} is more similar to our tests on synthetic applications. Their test on CPU and
network transfers have qualitatively the same result as the results we got. Their CPU tests show the same performance on both kind of virtualization
platforms and for the network transfer test of single client and single server they have shown that VMs consume more power than containers which is
in agreement with what we have seen in our results for synthetic applications. The memory test of Morabito~\cite{morabito} on the other hand show
that VMs consume almost the same power as containers while our results show that a little more power is consumed by VMs than containers.

Kritwara~et~al.~\cite{rattanaopas} investigated power consumption and performance issues concerning memory and disk I/O in Xen and KVM
virtualization environments. The power consumption and the latency in the disk I/O test results of Kritwara~et~al.~\cite{rattanaopas} are
reported to be decreasing function of the block size similar to what we obtained. The memory test in Kritwara~et~al.~\cite{rattanaopas} compares the
memory usage of \textit{Mysqlslap} queries in KVM and Xen which can not be compared with our memory test which allocates and deallocates a certain
amount of memory. Avino~et~al.~\cite{giuseppe} addressed the suitability of Docker in MEC scenarios by quantifying the CPU consumed by Docker when
running two different containerized services: multiplayer gaming and video streaming. They have done tests by varying the number of clients and
servers for both services. For the gaming service, the overhead logged by Docker increased only with the number of servers; conversely, for the video
streaming case, the overhead is not affected by the number of either clients or servers. The work by Avino~et~al.~\cite{giuseppe} has followed similar
methodology in observing system resource utilization of real-world applications except that they do not present power consumption measurements. We see
that the CPU utilization has the same qualitative behavior.

Fan~et~al.~\cite{qiang} proposed the Energy driven AvataR migration (EARN) scheme to reduce the total on-grid energy consumption of GCN by
considering the energy consumption of Avatar migrations.

\section{Conclusion}
\label{sec:conclusion}

We aimed to assess the  power consumption due to VM- and container-based virtualization. To this end, we performed an extensive set of real-world measurements, using VirtualBox and Docker as reference technologies and a wide set of workloads, both synthetic and real-world.

Through our measurements, we found that CPU usage is the main driver of the global power consumption, and the extra power consumption of container-based virtualization is not only lower than that of VM-based virtualization, but also grows more slowly as the workload increases. Additionally, we observed that that there is a penalty in energy consumption associated with creating a VM and then leaving it unused, while containers only consume system resources if the application they host is active.
These findings suggest that container-based virtualization is an attractive technology for MEC scenarios, when large numbers of virtualized applications run on the same physical hardware and some of them may be inactive for some time periods, e.g., during user handover.

\bibliography{refs}%

\begin{thebibliography}{10}

\bibitem{fog}
Bonomi Flavio, Milito Rodolfo, Zhu Jiang, Addepalli Sateesh. Fog computing and
  its role in the internet of things.  In: ; 2012.

\bibitem{mec}
Taleb Tarik, Samdanis Konstantinos, Mada Badr, Flinck Hannu, Dutta Sunny,
  Sabella Dario. On multi-access edge computing: A survey of the emerging 5G
  network edge cloud architecture and orchestration.  {\it IEEE Communications
  Surveys \& Tutorials. }2017;.

\bibitem{vms}
Rosenblum Mendel, Garfinkel Tal. Virtual machine monitors: Current technology
  and future trends.  {\it IEEE Computer. }2005;.

\bibitem{containers}
Felter Wes, Ferreira Alexandre, Rajamony Ram, Rubio Juan. An updated
  performance comparison of virtual machines and linux containers.  In: ; 2015.

\bibitem{ols}
Marquardt Donald~W. An algorithm for least-squares estimation of nonlinear
  parameters.  {\it Journal of the society for Industrial and Applied
  Mathematics. }1963;.

\bibitem{kansal}
Kansal Aman, Zhao Feng, Liu Jie, Kothari Nupur, Bhattacharya Arka~A. Virtual
  machine power metering and provisioning.  In: ; 2010.

\bibitem{krishnan}
Krishnan Bhavani, Amur Hrishikesh, Gavrilovska Ada, Schwan Karsten. {VM power
  metering: feasibility and challenges}.  {\it ACM SIGMETRICS Performance
  Evaluation Review. }2011;.

\bibitem{ata}
Ata E, Bohra Husain, Chaudhary Vipin. VMeter: Power Modelling for Virtualized
  Clouds.  In: ; 2010.

\bibitem{maxime}
Colmant Maxime, Kurpicz Mascha, Felber Pascal, Huertas Loic, Rouvoy Romain,
  Sobe Anita. Process-level Power Estimation in VM-based Systems.  In: ; 2015.

\bibitem{gaurav}
Dhiman Gaurav, Mihic Kresimir, Rosing Tajana. A System for Online Power
  Prediction in Virtualized Environments Using Gaussian Mixture Models.  In: ;
  2010.

\bibitem{bertran}
Bertran Ramon, Becerra Yolanda, Carrera David, et al. Accurate energy
  accounting for shared virtualized environments using pmc-based power modeling
  techniques.  In: ; 2010.

\bibitem{morabito}
Morabito Roberto. Power consumption of virtualization technologies: an
  empirical investigation.  In: ; 2015.

\bibitem{rattanaopas}
Kritwara Rattanaopas, Tandayya Pichaya. Comparison of disk I/O power
  consumption in modern virtualization.  In: ; 2015.

\bibitem{giuseppe}
Avino Giuseppe, Malinverno Marco, Malandrino Francesco, Casetti Claudio,
  Chiasserini Carla-Fabiana. Characterizing Docker Overhead in Mobile Edge
  Computing Scenarios.  In: ; 2016.

\bibitem{qiang}
Fan Qiang, Ansari Nirwan, Sun Xiang. Energy Driven Avatar Migration in Green
  Cloudlet Networks.  In: ; 2016.

\end{thebibliography}

\section*{Acknowledgement}
This work is supported by the European Commission through the H2020 5G-TRANSFORMER project (Project ID 761536).

\end{document}